\documentclass[sigconf]{acmart}
\renewcommand\footnotetextcopyrightpermission[1]{}
\AtBeginDocument{%
  \providecommand\BibTeX{{%
    \normalfont B\kern-0.5em{\scshape i\kern-0.25em b}\kern-0.8em\TeX}}}

\usepackage{longtable}
\usepackage{listings}
\usepackage{hyperref}
\usepackage{makecell}
\usepackage{multirow}
\usepackage{tikz}
\usetikzlibrary{snakes}
\usetikzlibrary{decorations.pathmorphing}
\usepackage{dblfloatfix}
\usepackage{afterpage}

\newcommand\mynode[2][]{\tikz\node (#1) {#2};}
\tikzstyle{every picture}+=[remember picture,baseline]

\setcopyright{none}
\copyrightyear{2022}
\acmYear{2022}
\acmDOI{XXXXXXX.XXXXXXX}

\acmConference[HEART '22]{12th International Symposium on Highly Efficient Accelerators and Reconfigurable Technologies}{June 09--10,
  2022}{Tsukuba, Japan}
\acmPrice{15.00}
\acmISBN{978-1-4503-XXXX-X/18/06}

\begin{document}

\title[Low-power option Greeks: Efficiency-driven market risk analysis using FPGAs]{Low-power option Greeks: \\ Efficiency-driven market risk analysis using FPGAs}

\author{Mark Klaisoongnoen}
\email{Mark.Klaisoongnoen@ed.ac.uk}
\affiliation{%
  \institution{EPCC at the University of Edinburgh}
  \streetaddress{The Bayes Centre}
  \city{Edinburgh}
  \country{UK}
}

\author{Nick Brown}
\affiliation{%
  \institution{EPCC at the University of Edinburgh}
  \streetaddress{The Bayes Centre}
  \city{Edinburgh}
  \country{UK}
}

\author{Oliver Thomson Brown}
\affiliation{%
  \institution{EPCC at the University of Edinburgh}
  \streetaddress{The Bayes Centre}
  \city{Edinburgh}
  \country{UK}
}

\renewcommand{\shortauthors}{Klaisoongnoen, et al.}

\begin{abstract}
Quantitative finance is the use of mathematical models to analyse financial markets and securities. Typically requiring significant amounts of computation, an important question is the role that novel architectures can play in accelerating these models. In this paper we explore the acceleration of the industry standard Securities Technology Analysis Center's (STAC) derivatives risk analysis benchmark STAC-A2\texttrademark{} by porting the Heston stochastic volatility model and Longstaff and Schwartz path reduction onto a Xilinx Alveo U280 FPGA with a focus on efficiency-driven computing.

Describing in detail the steps undertaken to optimise the algorithm for the FPGA, we then leverage the flexibility provided by the reconfigurable architecture to explore choices around numerical precision and representation. Insights gained are then exploited in our final performance and energy measurements, where for the efficiency improvement metric we achieve between an 8 times and 185 times improvement on the FPGA compared to two 24-core Intel Xeon Platinum CPUs. The result of this work is not only a show-case for the market risk analysis workload on FPGAs, but furthermore a set of efficiency driven techniques and lessons learnt that can be applied to quantitative finance and computational workloads on reconfigurable architectures more generally.
\end{abstract}


\maketitle

\section{Introduction}



Market risk analysis involves determining the impact of price movements on financial positions held by investors or traders. Sitting under the broader field of quantitative finance, the use of mathematical models and datasets to analyse financial markets, such workloads are heavy users of computational resource. Whilst running these models on CPUs is currently dominant, there have been some successes with exploring the acceleration of quantitative finance using FPGAs \cite{brown2021optimisation} \cite{inggs2015high} \cite{diamantopoulos2021acceleration}. However to date the majority use of FPGAs in the financial world has been in high-frequency trading.

One of the blockers to FPGAs gaining traction in quantitative finance is the historically significant time investment required in programming reconfigurable architectures and need for detailed hardware-level knowledge on behalf of developers. Whilst the benefits of FPGAs to high-frequency trading have been sufficiently obvious to make programming FPGAs in hardware description languages (HDL) such as VHDL or Verilog worthwhile, the benefits have been less forthcoming for quantitative workloads to warrant such efforts. Nevertheless in recent years FPGAs have become far more capable both in terms of hardware and software development ecosystem, and with tool chains such as Xilinx' Vitis \cite{vitis_exerpience_report}, one can now program FPGAs by writing code in C or C++. Consequently the increased programmability of these devices means that programming an FPGA is now much more a question of software development rather than hardware design, and this has been a major enabler for numerous communities to recently explore FPGAs for their workloads \cite{brown2021porting} \cite{yang2019fully} \cite{brown2021accelerating} more in-depth. 

Quantitative finance is one of these communities interested in the potential performance and energy advantages of FPGAs, and in this paper we explore porting models comprising a major component of the STAC-A2 market risk analysis benchmark to an Alveo U280 FPGA. The paper is structured as follows; in Section \ref{sec:bg} we briefly survey related activities and describe the context of this work, before in Section \ref{sec:experiment} detailing the experimental setup used throughout this paper and report baseline performance and energy of our benchmark kernel of interest on the CPU across numerous problem sizes. Section \ref{sec:optimisation} then describes the porting and optimisation of the code from the Von Neumann based CPU algorithm to a dataflow representation optimised for the FPGA, before exploring the performance and energy impact of changing numerical representation and precision. In Section \ref{sec:performance} we report multi-kernel performance and energy usage based upon applying the optimisation techniques and appropriate numerical choices that were highlighted in the previous section. This paper then concludes in Section \ref{sec:conclusions} before describing further work. The result of this work is not only a comprehensive efficiency-driven exploration of major components of STAC-A2 on the Alveo FPGA, but furthermore lessons that can be applied more widely to high performance numerical modelling on FPGAs.

\section{Background and related work}
\label{sec:bg}

The ability for FPGAs to provide low latency handling of data has meant that they have been successfully applied to high-frequency trading for a number of years \cite{leber2011high}. Traditionally such high-frequency codes were written in HDL \cite{aldridge2013high}, however FPGAs are yet to gain ubiquity in quantitative finance for accelerating financial computational models, and a major reason is that their benefits have been less clear. In recent years vendors such as Xilinx and Intel have invested significantly in new generations of more capable hardware and substantially improved their software development ecosystems. Xilinx's Vitis toolchain \cite{vitis} is an example, where using High Level Synthesis (HLS) programmers can write code for FPGAs in C or C++. This technology significantly improves productivity, opening up the programming of FPGAs to a much wider community, and significantly reducing the barrier to entry. Consequently these advantages makes the use of FPGAs more realistic for computational workloads such as quantitative finance, enabling software developers to port their codes more easily.

Nevertheless HLS is not a silver bullet, and whilst this technology has made the physical act of programming FPGAs much easier, one must still select appropriate kernels that will suit execution on FPGAs \cite{brown2020exploring} and recast their Von Neumann style CPU algorithms into a dataflow style \cite{koch2016fpga} to obtain best performance. Whilst there have been some successes in accelerating quantitative finance on FPGAs \cite{brown2021optimisation} \cite{inggs2015high} \cite{diamantopoulos2021acceleration}, and Xilinx have recently provided support in their open source Vitis Library \cite{vitis_libraries} for numerous quantitative finance primitives, there is still much exploration to be undertaken especially with the objective of efficiency-driven computing looking to optimise both performance and energy efficiency. Furthermore, more development is needed of the underlying algorithmic techniques to inform software developers how best to port their codes to FPGAs.

\subsection{Securities Technology Analysis Center}
The Securities Technology Analysis Center (STAC) acts as a forum for some of the world's largest global banks, hedge funds and hardware companies in the area of finance. With membership comprising over 400 financial institutions and more than 50 technology vendors, STAC provide industry standard financial benchmarks suites representing common workloads. Based on STAC's benchmark specifications, members can test, optimise and validate their technology against these world-leading benchmarks to compare their software codes and hardware infrastructure against a common market baseline. When undertaking such audits STAC members must comply with strict rules, and while this is beneficial for a fair comparison, in this research we are using the benchmarks differently as we are not looking to undertake any official audits and results should not be compared to audited results. Instead, we use selected benchmarks as drivers to explore algorithmic, performance, and energy properties of FPGAs, consequently meaning that we are able to leverage components of the benchmarks in a more experimental manner. 

\subsection{STAC-A2: Market risk analysis}
\label{sec:staca2}
The STAC-A2 benchmark \cite{staca2} focuses on real-world market risk analysis \cite{stac-a2-2012} which is an important, ongoing task for investors, trading firms and regulatory authorities. Financial models are deployed to analyse the impact of price movements in the market on financial positions held by investors. Understanding the risk carried by individual or combined positions is crucial for such organisations, and provides insights how to adapt trading strategies into more risk tolerant or risk averse positions. The quality of market risk management is not only driven by demand from investors to track changing market conditions, but also due to increased requirements by regulatory authorities. With expanding numbers of financial positions in a portfolio and increasing market volatility, the complexity and workload of risk analysis has risen substantially in recent years and requires model computations that yield insights for trading desks within acceptable time frames.

Market risk analysis relies on analysing financial derivatives which derive their value from an underlying asset, such as a stock, where an asset's price movements will change the value of the derivative. For each asset, risk analysis relies on understanding sensitivities to market changes which is known as Greeks. Computing these risk sensitivity Greeks involves a high computational workload based on numerical models and many financial firms manage dedicated data centres which are, in-part, apportioned to this workload. Whilst computational performance is one essential aspect of effective risk analysis, energy efficiency is also important because of the dedicated infrastructure in-place and increased frequency of generation and usage of derived risk information in general.

It is therefore worthwhile exploring opportunities for efficiency-driven market risk analysis by leveraging reconfigurable architectures, benefiting from their innate energy efficiency, and the industry standard STAC-A2 benchmark. The benchmark itself involves path generation for each asset using the Andersen Quadratic Exponential (QE) method \cite{andersen2007efficient} which undertakes time-discretization and Monte Carlo simulation of the Heston stochastic volatility model \cite{closed-form-heston-stochastic-model} before pricing the option using Longstaff and Schwartz \cite{longstaff-schwartz} for early option exercise. Previously Xilinx developed a proprietary implementation of this benchmark on their Alveo U250 FPGAs which, when running over eight U250s, obtained a 1.48 times speed up compared to the CPU \cite{xilinx-staca2} in an official STAC audit. 

As an official STAC audit Xilinx had to comply with a strict series of guidelines that ensure results are highly trustworthy, but these govern what can and can not be changed in the code, and therefore limit flexibility. By contrast we are not undertaking an official audit, but instead using the benchmark as a vehicle to better understand the use of reconfigurable architectures for this workload and develop appropriate dataflow techniques. Consequently we have more choice around which parts we offload and are able to undertake more extensive code level changes. 

When profiling STAC-A2 on the CPU we found that over 97\% of the runtime for the reference implementation was spent on the Heston stochastic volatility model and path reduction in Longstaff and Schwartz. However only around 50\% of the CPU cycles were completing useful work in these parts of the code, with approximately 20\% of cycles stalled due to memory bottlenecks and the rest stalled due to other core-bound issues. Consequently an important question is whether, by exploiting the ability of reconfigurable architectures to tailor the electronics to the code, it is possible to ameliorate these CPU issues and at the same time benefit from the typically greater energy efficiency of the FPGA \cite{betkaoui2010comparing}.

\subsubsection{General code structure}

The components of the STAC-A2 benchmark we are focusing on in this work operate over paths, which can be thought of as the accuracy of the sensitivities that are being computed. For each path the benchmark works across assets and timesteps, the former representing distinct derivatives that sensitivities are being computed for and the later denotes time with one timestep per trading day. Each kernel of the code typically loops through in this order, paths as the outer, assets as the middle and timesteps as the inner loops.

Each asset has an associated Heston model configuration and this is used as input along with two double precision numbers for each path, asset, and timestep to calculate the variance and log price for each path and follow Andersen's QE method \cite{andersen2007efficient}. Subsequently the exponential of the result for each path of every asset of every timestep is computed. Results from these calculations are then used an an input to the Longstaff and Schwartz model. The Longstaff and Schwartz model comprises two parts, a reduction calculation and then a quadratic curve fit. We only undertake the reduction on the FPGA because this accounts for the majority of the runtime in that model and code required as part of quadratic curve fit is more verbose and less suited to reconfigurable architectures. All computations in the reference implementation are undertaken, by default, using double precision floating-point arithmetic, and in total there are 307 floating-point arithmetic operations required for each element (every path of every asset of every timestep).

\section{Experiment and benchmark setup}
\label{sec:experiment}
Table \ref{tab:prob_sizes_defined} defines the five classes of problem size used throughout this work in evaluating the CPU and FPGA benchmark implementations. It should be stressed that these problem sizes do not represent an official STAC audit configuration, but instead have been selected in this research to provide a wide range of data sizes under test. For these problem sizes we vary the number of timesteps, ranging from 6 months for the tiny problem size to 5 years for the huge problem size, and assets being studied. As described in Section \ref{sec:staca2}, each element comprises two double precision numbers hence in Table \ref{tab:prob_sizes_defined} the number of data points is double the number of elements. All experiments undertaken report results from executing the Heston stochastic volatility model and path reduction in Longstaff and Schwartz which are our areas of focus in this work.

\begin{table}[h]
  \caption{Problem Sizes with defined number of assets ($A$), time steps ($T$) and paths ($P$). The number of elements is the product of $A * T * P$. Each element requires two data points (Dpoints). Each data point is a 64-bit floating-point number.}
  \label{tab:prob_sizes_defined}
  \begin{tabular}{c|ccc|ccc}
    \toprule
    \makecell{\textbf{Problem} \\ \textbf{size}} & \makecell{\textbf{A}} & \makecell{\textbf{T}} & \makecell{\textbf{P}} & \makecell{Elements \\ $\left(\times 10^6\right)$} & \makecell{Dpoints \\ $\left(\times 10^6\right)$} & \makecell{Size \\ (MB)} \\
    \midrule
    Tiny (T)         & 5  & 126 & 25 k  & 15.75 & 31.5 & 252 \\
    Small (S)        & 10  & 126 & 25 k  & 31.5 & 63 & 504 \\
    Medium (M)       & 20 & 252 & 25 k  & 126 & 252 & 2016 \\
    Large (L)        & 30 & 504 & 25 k  & 378 & 756 & 6048 \\
    Huge (H)         & 50 & 1260 & 25 k & 1575 & 3150 & 25200 \\
  \bottomrule
\end{tabular}
\end{table}

All CPU runs are undertaken by threading, using OpenMP, across two 24-core Xeon Platinum (Cascade Lake) 8260M CPUs which are fitted into a single node of our test system and energy measured via RAPL. All CPU runs are executed across all 48 physical cores as this was found to be the optimal CPU configuration. For the FPGA runs we use a Xilinx Alveo U280, running at the default clock frequency of 300MHz, which contains an FPGA chip with 1.08 million LUTs, 4.5MB of on-chip BRAM, 30MB of on-chip UltraRAM, and 9024 DSP slices. This PCIe card also contains 8GB of High Bandwidth Memory (HBM2) and 32GB of DDR DRAM on the board. The FPGA card is hosted in a system with a 26-core Xeon Platinum (Skylake) 8170 CPU and energy metrics gathered via the Xilinx Runtime Library (XRT). All bitstreams are built using the Xilinx Vitis framework version 2021.2 which at the time of writing is the latest version. All reported results are averaged over five runs and total FPGA runtime and energy usage includes measurements of the kernel, data transfer and any required data reordering on the host.

\subsection{CPU performance and energy baseline}
Table \ref{tab:cpu_baseline_shortenedtab} reports performance and energy usage of the STAC-A2 Heston stochastic volatility model and Longstaff and Schwartz path reduction running over the two 24-core Xeon Platinum CPUs across the problem sizes described in Table \ref{tab:prob_sizes_defined} \footnotemark. Specifically it is these performance and energy values that we are aiming to improve upon by porting to the FPGA in this work.

\footnotetext{The experiments conducted have not been designed to comply with official STAC benchmarking rules and regulations. Therefore the experimental results that we present are of a research nature and are not representative of official STAC audits.}

\begin{table}[h]
  \caption{CPU baseline performance on two 24-core Intel Xeon Platinum 8260M CPUs across all 48 physical cores.} 
  \label{tab:cpu_baseline_shortenedtab}
  \begin{tabular}{c|cc|cc}
    \toprule
    & \multicolumn{2}{c|}{\textbf{Single precision}} & \multicolumn{2}{c}{\textbf{Double precision}} \\
    \midrule
    \makecell{\textbf{Problem} \\ \textbf{size}} & \makecell{Runtime\\ (ms)}  & \makecell{Energy\\ (J)} & \makecell{Runtime\\ (ms)} & \makecell{Energy\\ (J)} \\ 
    \midrule
     Tiny (T)      &   372.15  &   93.72  &    369.07  &   94.35 \\
     Small (S)     &   629.18  &  151.50  &    638.67  &  153.85 \\
     Medium (M)    &  1545.94  &  397.48  &   1551.90  &  393.34 \\
     Large (L)     &  4574.41  & 1277.14  &   4558.11  & 1266.59 \\
     Huge (H)      & 15825.42  & 5011.16  &  15561.09  & 4900.16 \\
  \bottomrule 
  \end{tabular}
\end{table}

\section{FPGA porting and optimisation techniques}
\label{sec:optimisation}
\subsection{Algorithmic optimisations}
\label{sec:algopt}
As described in Section \ref{sec:staca2}, in this work we focus on porting the STAC-A2 Heston model and Longstaff and Schwartz path reduction functionality onto an Alveo U280 FPGA. Table \ref{tab:fpga_versions} reports performance, card power (average power drawn by FPGA card only), and total energy (energy used by FPGA card and host for data manipulation) for different versions of a single FPGA kernel implementing these models for the tiny benchmark size and against the two 24-core CPUs for comparison. In this subsection we focus mainly on performance, with the objective being that by reducing the runtime this will then also help to reduce the total energy used. The total runtime, which includes data transfer to and from the FPGA, and kernel-only runtime are reported, with \emph{initial FPGA} being the first version of the kernel on the FPGA. At around 62 times slower than the CPU, this first version left plenty of opportunity for optimisation, largely because it was still Von Neumann based, and additionally it also drew the largest average card power of all the FPGA versions, although the power differences are fairly minimal but this combination resulted in significant energy usage.

\begin{table}
    \caption{Single FPGA kernel performance of tiny problem size for different versions and against Xeon Platinum CPUs}
  \label{tab:fpga_versions}
  \begin{tabular}{ccccc}
    \toprule
    \textbf{Description} & \makecell{\textbf{Total} \\ \textbf{Runtime} \\ \textbf{(ms)}} & \makecell{\textbf{Kernel}\\ \textbf{Runtime} \\ \textbf{(ms)}} & \makecell{\textbf{Card} \\ \textbf{Power} \\ \textbf{(W)}} & \makecell{\textbf{Total}\\\textbf{Energy}\\ \textbf{(J)}}\\ 
    \hline
    Two 24-core CPUs & 369.07 & - & - & 94.35\\
    Initial FPGA & 22882.23 & 22829.36 & 30.58 & 699.73\\
    Dataflow enabled & 9307.98 & 9267.43 & 29.87 & 278.03\\
    Loop interchange & 236.39 & 179.83 & 29.34 & 9.12\\
    Double buffering & 115.35 & 72.32 & 29.67 & 5.51\\
  \bottomrule
\end{tabular}
\end{table}

\begin{figure}[h]
  \centering
  \includegraphics[scale=0.53]{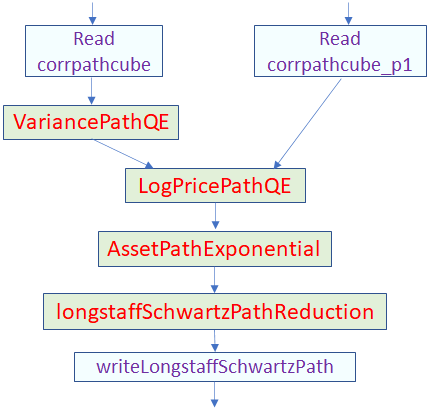}
  \caption{Illustration of dataflow design where stages are running concurrently connected via HLS streams}
  \label{fig:dataflowdesign}
\end{figure}

To optimise the kernel for FPGAs we first refactored top-down, moving the code into a dataflow style. This is illustrated in Figure \ref{fig:dataflowdesign} where the algorithm was decomposed into constituent components each of which is a separate function called from within an HLS \emph{DATAFLOW} region. These functions are running concurrently and connected by HLS streams, meaning that data is capable of continually streaming from one to the next with progress made each cycle. External memory accesses were also batched into width of 512-bits as per best-practice \cite{hls_best_practice}. The performance of adopting this dataflow approach is reported by \emph{dataflow enabled} in Table \ref{tab:fpga_versions}, and whilst it can be seen this significantly improved performance, it was still 25 times slower than the CPU.

The reason for this performance shortfall was that, whilst we had optimised top-down by splitting the kernel into concurrently running dataflow regions, within each region the code was frequently stalling. Put simply, we were failing to \emph{keep the data flowing} because we had not yet also optimised bottom-up at the individual loop level. Most egregious was the fact that numerous spatial dependencies were limiting effective pipelining of loops and this is illustrated in Listing \ref{lst:spatial_dependency} for the \emph{logPricePathQE} function. It can be seen that there are three loops with the inner, \emph{timesteps} loop, calling the \emph{Y1QE} function for each iteration and results written to the \emph{i+1} element of the \emph{asspath} array. However an input to the \emph{Y1QE} function is the resulting value calculated at the previous timestep loop iteration, \emph{asspath[i]}, and consequently the call to \emph{Y1QE} in one iteration depends upon results calculated at the previous iteration. This was problematic because \emph{Y1QE} undertakes 37 double precision floating-point operations which in total requires 457 cycles, and all these cycles must have completed before the next inner loop iteration can start to be processed. 

\begin{lstlisting}[frame=lines,caption={Illustration of spatial dependency issue},label={lst:spatial_dependency}, numbers=left]
void logPricePathQE(unsigned int timesteps, ..., double* asspath) {
  for (unsigned int j=0; j<paths; j++) {
    for (unsigned int k=0; k<assets; k++) {
      asspath[0] = .....;
      for (unsigned int i=0; i<timesteps; i++) {
#pragma HLS PIPELINE II=1	
        double W = ....;
        asspath[i+1] = Y1QE(..., asspath[i], W);
} } } }
\end{lstlisting}

To optimise performance one should aim for the loop level initiation interval, the number of cycles between processing one iteration and the next, to be one \cite{hls_best_practice}, where a new loop iteration is processed each cycle. Due to the spatial dependency a new iteration could only be processed every 457 cycles and the reason was that, based on the mathematics of the algorithm, there is a dependency between the processing of one timestep and the next. We therefore needed to ensure that there would be at-least 457 cycles between processing subsequent timesteps, and loop interchange was undertaken to achieve this, moving the outer loop over \emph{paths} to be the inner loop. 

This loop interchange is sketched in Listing \ref{lst:spatial_dependency_fixed}, where the results for each path's \emph{Y1QE} calculation is cached in \emph{cached\_asspath} and whilst the calculations involving one timestep still depend on the previous timestep for that path, by undertaking this reordering there are \emph{paths} cycles between each subsequent timestep iteration (i.e. each middle loop iteration). As long as the number of paths is greater than 457 then there is no longer a spatial dependency. The HLS dependence pragma at line 3 in Listing \ref{lst:spatial_dependency_fixed} is required because the number of paths is a runtime parameter and consequently HLS can not guarantee at synthesis time that this is large enough.

\begin{figure}[h]
  \centering
  \includegraphics[scale=0.27]{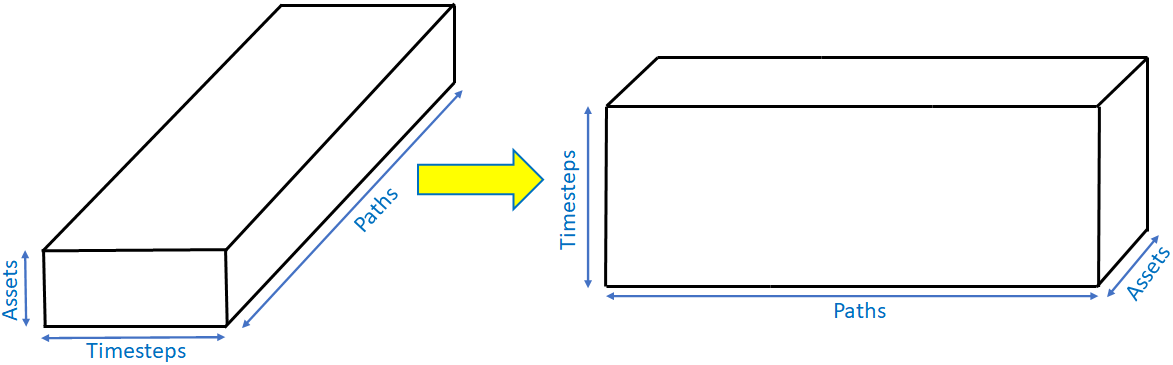}
  \caption{Change to data layout required for loop interchange}
  \label{fig:datalayout}
\end{figure}

Listing \ref{lst:spatial_dependency_fixed} sketches this loop reordering for the \emph{logPricePathQE} dataflow region, and this was also needed for the \emph{variancePathQE} region for similar reasons. Facilitating this reordering required changing the data layout, as illustrated in Figure \ref{fig:datalayout}, however doing so resulted in data streamed from \emph{AssetPathExponential} in Figure \ref{fig:dataflowdesign} in the wrong orientation for the subsequent \emph{longstaffSchwartzPathReduction} calculation. This Longstaff Schwartz function undertakes a reduction across assets, calculating the maximum value for each path and timestep held by any asset. In the \emph{longstaffSchwartzPathReduction} function we therefore create a local on-chip buffer of size paths by timesteps, which acts as a local cache to store the current maximum value for each asset and whose values are read and updated as data streams in. However, this caching approach limits the number of paths and timesteps to the amount of on-chip memory, for instance the tiny problem size would require around 25MB of on-chip memory and the huge problem size around 250MB. These sizes are beyond the on-chip BRAM memory available on the Alveo U280 and consequently we decompose paths into batches, within each batch looping over the assets, timesteps and number of paths in that batch before moving onto the next batch. 

\begin{lstlisting}[frame=lines,caption={Reordering data and loops to fix spatial dependency},label={lst:spatial_dependency_fixed}, numbers=left]
void logPricePathQE(unsigned int timesteps, ..., double* asspath) {
  double cached_asspath[MAX_PATHS];
#pragma HLS dependence variable=cached_asspath inter false
  for (unsigned int k=0; k<assets; k++) {
    for (unsigned int i=0; i<timesteps; i++) {
      for (unsigned int path=0; path<paths; path++) {
#pragma HLS PIPELINE II=1
        double W = ....;
        asspath=Y1QE(..., cached_asspath[path], W);
        cached_asspath[path]=asspath;
} } } }
\end{lstlisting}

\begin{figure}[h]
  \centering
  \includegraphics[scale=0.33]{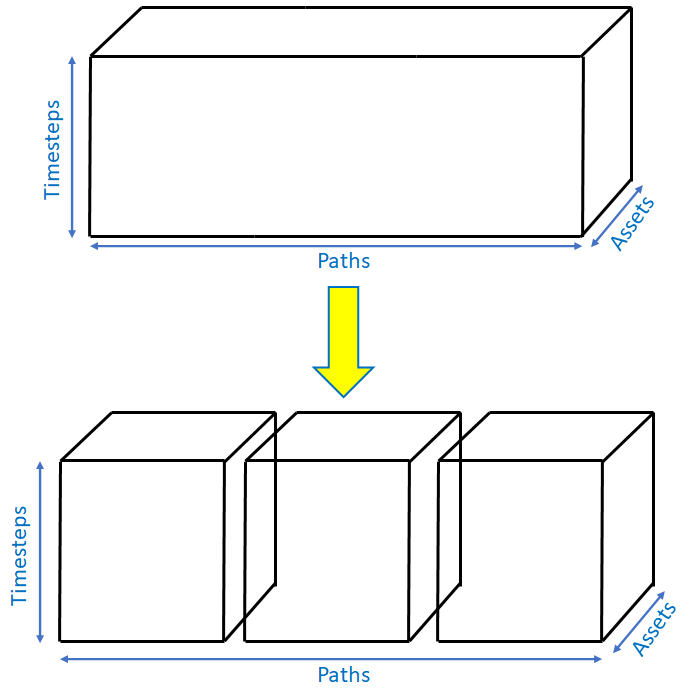}
  \caption{Illustration of data decomposition along paths dimension into batches that are then processed consecutively}
  \label{fig:decompose}
\end{figure}

This decomposition of paths into batches is illustrated in Figure \ref{fig:decompose}, where it can be seen that the \emph{paths} dimension has been split in this illustration into three batches. Each batch is processed completely before the next is started, and as long as the number of paths in each batch is greater than 457, the depth of the pipeline in \emph{Y1QE}, then calculations can still be effectively pipelined. The on-chip memory required for caching in the \emph{longstaffSchwartzPathReduction} calculation is still fairly large, around 5MB for path batches of size 500 paths and 1260 timesteps, and therefore we place this in the Alveo's UltraRAM rather than smaller BRAM. 

Ultimately the objective had been to optimise the loops bottom-up, but doing so required a reorganisation of the data which then had a knock on effect of requiring caching and batching of paths to fit the memory of the architecture. The performance of our kernel on the Alveo U280 at this point is reported by \emph{loop interchange} in Table \ref{tab:fpga_versions}, where we are working in batches of 500 paths per batch, and hence 50 batches, and it can be observed that the FPGA kernel is now outperforming the two 24-core Xeon Platinum CPUs for the first time. It can also be seen that the difference between the total and kernel execution times is greater than with previous versions, this is because now total runtime also includes the time required to reorder input data on the host before transfer, and reordering of result data after kernel execution because of the revised data layout depicted in Figure \ref{fig:datalayout}. Likewise energy for these reordering activities is also included as part of total energy in Table \ref{tab:fpga_versions}.

\begin{figure}[h]
  \centering
  \includegraphics[scale=0.53]{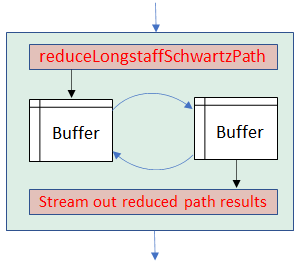}
  \caption{Illustration of double buffering approach over batches of path for longstaffSchwartzPathReduction}
  \label{fig:doublebuffer}
\end{figure}

\begin{table*}[!t]
  \caption{Percentage deviation (lower is better) for each data type running on the FPGA from the reference implementation running in double precision floating-point on the CPU. The deviation is independent from the number of assets or paths. }
  \label{tab:dtypes_accuracy}
  \begin{tabular}{c|ccc|cccccccc}
    \toprule

    \makecell{\textbf{timesteps}} & \multicolumn{3}{c|}{\textbf{floating-point}} & \multicolumn{8}{c}{\textbf{ap\_fixed}} \\

    & half & float & double & <8,3> & <8,4> & <16,6> & <16,8> & <32,12> & <32,16> & <64,12> & <64,24> \\
    \midrule
    126 & 0.06\% & < 0.00\% & < 0.00\% & 326.80\% & 394.44\% & 112.44\% & 28.32\% & 1.39\% & 0.38\% & 100.00\% & 100.00\%\\
    252 & 0.38\% & < 0.00\% & < 0.00\% & 39.96\% & 373.43\% & 1887.67\% & 8514.76\% & 1.39\% & 0.83\% & 100.00\% & 100.00\%\\ 
    504 & 1.01\% & < 0.00\% & < 0.00\% & 27.78\% & 433.62\% & 2055.00\% & 6173.76\% & 1.44\% & 3.96\% & 100.00\% & 100.00\%\\ 
    1260 & 11.14\% & < 0.00\% & < 0.00\% & 38.17\% & 404.17\% & 286.26\% & 98.56\% & 1.75\% & 16.88\% & 100.00\% & 100.00\%\\ 
    2520 & 14.65\% & < 0.00\% & < 0.00\% & 25.43\% & 405.35\% & 1246.68\% & 98.29\% & 80.08\% & 11.84\% & 100.00\% & 100.00\%\\ 
  \bottomrule
\end{tabular}
\end{table*}

However the \emph{longstaffSchwartzPathReduction} function was filling the maximum values over assets in each batch of paths and only streaming out resulting maximum values from the cache once complete. This stalled the dataflow because, whilst filling was occurring, then no streaming was being performed and vice-versa. Consequently we adopted a double buffering approach, where the first buffer is filled for the current batch of paths and results from processing the previous batch are concurrently streamed out from the second buffer. This is illustrated in Figure \ref{fig:doublebuffer}, where \emph{buffer} is a ping-pong buffer that is switched between the two dataflow regions between each batch of paths. This enables the reduction calculation and streaming of output data to run concurrently from the second batch of paths onwards and Table \ref{tab:fpga_versions} reports this optimisation as \emph{double buffering}. It can be seen that the overall execution time (including data transfer and data reordering on the host) is now 3.2 times less than the two 24-core Xeon Platinum CPUs, and the kernel runtime alone (ignoring data transfer and data reordering) is 5.1 times less than the CPUs. It is noteworthy that our optimised kernel execution time on the FPGA is around 320 times faster than the initial, non-optimised, FPGA kernel execution time and this demonstrates the importance of applying such dataflow techniques to the algorithm. Interestingly these optimisations did not increase the power draw, and this combined with the significantly reduced runtime has resulted in approximately a 140 times reduction in energy draw between the initial and the optimised FPGA versions, and requires 17 times less energy than the two CPUs.

\subsection{Numeric optimisation}
\label{sec:numeric}

The STAC-A2 benchmark specification requires double precision floating-point arithmetic which is the numerical representation used until this point. However the ability to tailor execution on the FPGA means provides more flexibility than on the CPU, where Xilinx's Vitis HLS supports double, single, and half precision floating-point data types as well as arbitrary precision fixed-point. Consequently it is instructive to explore the properties of performance, power draw, energy efficiency, accuracy, and resource utilisation for these alternative numerical precision and representations. 

\begin{table}
  \caption{Resource utilisation on the U280 for a single FPGA kernel with chosen data types implemented on FPGA in percentage of the available resources. These are all built with 500 paths per batch and a maximum number of 1260 timesteps. Asterisk (*) marks fixed-point precision data types.}
  \label{tab:resource_utilisation}
  \begin{tabular}{c|cccccc}
    \toprule
    \makecell{\textbf{dtype}} 
    & \makecell{\textbf{LUT}} 
    & \makecell{\textbf{LUT}\\\textbf{Mem}} 
    & \makecell{\textbf{REG}} 
    & \makecell{\textbf{BRAM}} 
    & \makecell{\textbf{URAM}} 
    & \makecell{\textbf{DSP}} \\
    
    \midrule
    half & 1.99\% & 0.62\% & 1.49\% & 1.32\% & 16.04\% & 2.59\% \\
    float & 3.53\% & 0.88\% & 2.18\% & 1.60\% & 16.04\% & 3.53\% \\
    double & 10.11\% & 2.35\% & 5.77\% & 2.92\% & 16.04\% & 8.88\% \\
    <8,3>* & 2.96\% & 0.53\% & 1.79\% & 1.43\% & 16.04\% & 1.42\% \\
    <8,4>* & 2.81\% & 0.50\% & 1.73\% & 1.43\% & 16.04\% & 1.43\% \\
    <16,6>* & 7.02\% & 0.93\% & 3.71\% & 2.10\% & 16.04\% & 3.06\% \\
    <16,8>* & 7.02\% & 0.93\% & 3.71\% & 2.10\% & 16.04\% & 3.06\% \\
    <32,12>* & 11.17\% & 0.97\% & 5.21\% & 1.88\% & 16.04\% & 2.71\% \\
    <32,16>* & 21.46\% & 2.34\% & 9.95\% & 3.31\% & 16.04\% & 7.84\% \\
    <64,12>* & 1.70\% & 0.19\% & 0.69\% & 0.50\% & 16.04\% & 0.71\% \\
    <64,24>* & 1.69\% & 0.19\% & 0.68\% & 0.50\% & 16.04\% & 0.71\% \\
  \bottomrule
\end{tabular}
\end{table}

As the host CPU only supports double and single precision floating-point natively, data is transferred from host to device in either one of these formats (double precision for 64-bit fixed or floating-point, and single precision for all others) and then typecast to the respective data type implemented on the FPGA if required. Results are subsequently typecast back to to appropriate host floating-point precision before transferred from the device. Consequently computation on the FPGA is undertaken in our chosen precision and floating or fixed-point, but data transfer always occurs in double or single precision floating-point.

Table \ref{tab:dtypes_accuracy} presents an accuracy comparison when using these different data representations on the FPGA. Deviations can proliferate and hence we explore how the accuracy changes as the number of timesteps progresses. Table \ref{tab:dtypes_accuracy} reports percentage accuracy against results obtained with the reference implementation running in double precision floating-point arithmetic on the CPU. It can be seen that on the FPGA floating-point double, single, and half precision, along with fixed-point single precision yield the lowest deviations whereas other configurations diverge significantly.

Whilst the accuracy results reported in Table \ref{tab:dtypes_accuracy} indicate that some configurations would be unsuited for this benchmark, it is still instructive to explore their resource utilisation properties along with energy and performance. The FPGA's resource utilisation of these different configurations for a single FPGA kernel is reported in Table  \ref{tab:resource_utilisation}, where irrespective of the numerical representation the amount of UltraRAM (URAM) consumed is constant. This is because of the mapping imposed by the HLS tooling to fit the memory access pattern across the constituent URAM banks. Generally, double precision floating-point has around the highest general resource utilisation. Single and half precision floating-point are rather better, apart from the URAM, and there isn't a clear pattern of resource utilisation across the different fixed-point representations.

\begin{table*}[!t]
  \caption{Single kernel performance and power details for different numerical precision and representation across benchmark problem sizes. Asterisk (*) marks fixed-point precision data types. \textsuperscript{1} Input data is reordered on CPU before transfer to FPGA and result data is reordered on CPU after transfer from FPGA to CPU as described in Section \ref{sec:algopt}.}
  \label{tab:cpu_baseline}
  \begin{tabular}{c|cc|cccccc|ccc}

    \toprule
    \makecell{\textbf{Problem} \\ \textbf{size}} 
    & \makecell{\textbf{Dtype} \\ \textbf{on CPU}} 
    & \makecell{\textbf{Dtype} \\ \textbf{on FPGA}} 
    & \makecell{\textbf{Overall} \\ \textbf{(ms)}}
    & \makecell{\textbf{Reorder\textsuperscript{1}} \\ \textbf{(ms)}} 
    & \makecell{\textbf{Xfer on} \\ \textbf{(ms)}}   
    & \makecell{\textbf{Execute} \\ \textbf{(ms)}}    
    & \makecell{\textbf{Xfer off} \\ \textbf{(ms)}} 
    & \makecell{\textbf{Reorder\textsuperscript{1}} \\ \textbf{(ms)}} 
    & \makecell{\textbf{$\overline{Card}$} \\ \textbf{(W)}} 
    & \makecell{\textbf{$\overline{CPU}$\textsuperscript{1}} \\ \textbf{(W)}} 
    & \makecell{\textbf{$Total$} \\ \textbf{(J)}} \\

    \midrule
     Tiny (T)  & \mynode[T11] & half     &  105.97 & \mynode[T12] & \mynode[Txfer11]  & 78.25  & \mynode[Txfer12] & \mynode[T13]   & 29.06 & \mynode[T14] & 4.58 \\  
               &              & <16,6>*  &  104.38 &              &                   & 76.66  &                  &                & 28.38 &              & 4.47 \\
               &              & <16,8>*  &  107.54 &              &                   & 79.82  &                  &                & 28.66 &              & 4.59 \\  
               &              & float    &  102.85 &              &                   & 75.13  &                  &                & 29.01 &              & 4.48 \\
               &              & <32,12>* &  105.96 &              &                   & 78.24  &                  &                & 28.76 &              & 4.55 \\  
               & \mynode[T21] & <32,16>* &  107.59 & \mynode[T22] & \mynode[Txfer21]  & 79.87  & \mynode[Txfer22] & \mynode[T23]   & 29.99 & \mynode[T24] & 4.72 \\  
               & \mynode[T31] & double   &  115.35 & \mynode[T32] & \mynode[T33]      & 72.32  & \mynode[T34]     & \mynode[T35]   & 29.67 & \mynode[T36] & 5.51 \\
               &              & <64,12>* &  119.66 &              &                   & 76.63  &                  &                & 28.67 &              & 5.53 \\
               & \mynode[T41] & <64,24>* &  118.18 & \mynode[T42] & \mynode[T43]      & 75.15  & \mynode[T44]     & \mynode[T45]   & 28.58 & \mynode[T46] & 5.47 \\
    \midrule
    Small (S)  & \mynode[S11] & half     & 194.02 & \mynode[S12] & \mynode[Sxfer11] & 143.11 & \mynode[Sxfer12] & \mynode[S13]           & 28.62 & \mynode[S14] & 8.82 \\   
               &              & <16,6>*  & 191.18 &              &                  & 140.27 &                  &                        & 28.60 &              & 8.73 \\
               &              & <16,8>*  & 197.04 &              &                  & 146.13 &                  &                        & 28.38 &              & 8.86 \\
               &              & float    & 188.44 &              &                  & 137.53 &                  &                        & 28.97 &              & 8.72 \\   
               &              & <32,12>* & 194.05 &              &                  & 143.14 &                  &                        & 28.86 &              & 8.86 \\
               & \mynode[S21] & <32,16>* & 197.09 & \mynode[S22] & \mynode[Sxfer21] & 146.18 & \mynode[Sxfer22] & \mynode[S23]           & 29.33 & \mynode[S24] & 9.03 \\
               & \mynode[S31] & double   & 211.94 & \mynode[S32] & \mynode[S33]     & 132.35 & \mynode[S34]     & \mynode[S35]           & 29.50 & \mynode[S36] &10.12\\
               &              & <64,12>* & 219.87 &              &                  & 140.28 &                  &                        & 23.03 &              &10.04 \\
               & \mynode[S41] & <64,24>* & 217.17 & \mynode[S42] & \mynode[S43]     & 137.58 & \mynode[S44]     & \mynode[S45]           & 28.23 & \mynode[S46] &10.00 \\
    \midrule
    Medium (M) & \mynode[M11] & half     & 747.25 & \mynode[M12] & \mynode[Mxfer11] & 543.50 & \mynode[Mxfer12] & \mynode[M13] & 28.57 & \mynode[M14]   &36.00\\    
               &              & <16,6>*  & 736.40 &              &                  & 532.65 &                  &              & 28.64 &                &35.74\\
               &              & <16,8>*  & 758.64 &              &                  & 554.89 &                  &              & 28.35 &                &36.18\\
               &              & float    & 725.92 &              &                  & 522.17 &                  &              & 28.77 &                &35.52\\
               &              & <32,12>* & 747.27 &              &                  & 543.52 &                  &              & 28.84 &                &36.18\\
               & \mynode[M21] & <32,16>* & 758.57 & \mynode[M22] & \mynode[Mxfer21] & 554.82 & \mynode[Mxfer22] & \mynode[M23] & 29.45 & \mynode[M24]   &36.92\\   
               & \mynode[M31] & double   & 838.83 & \mynode[M32] & \mynode[M33]     & 502.55 & \mynode[M34]     & \mynode[M35] & 29.50 & \mynode[M36]   &45.90\\
               &              & <64,12>* & 868.92 &              &                  & 532.64 &                  &              & 28.04 &                &45.59\\
               & \mynode[M41] & <64,24>* & 858.54 & \mynode[M42] & \mynode[M43]     & 522.26 & \mynode[M44]     & \mynode[M45] & 28.15 & \mynode[M46]   &45.38\\

    \bottomrule
\end{tabular}
  \begin{tikzpicture}[overlay]
    \draw[|-|] (T11.north) -- (T21.south) node [rectangle,midway,fill=white,inner sep=1.2ex] {float};
    \draw[|-|] (T12.north) -- (T22.south) node [rectangle,midway,fill=white,inner sep=1.2ex] {8.54};
    \draw[|-|] (T13.north) -- (T23.south) node [rectangle,midway,fill=white,inner sep=1.2ex] {1.82};
    \draw[|-|] (T14.north) -- (T24.south) node [rectangle,midway,fill=white,inner sep=1.2ex] {173.75}; 
    \draw[|-|] (T31.north) -- (T41) node [rectangle,midway,fill=white,inner sep=1.2ex] {double}; 
    \draw[|-|] (T32.north) -- (T42) node [rectangle,midway,fill=white,inner sep=1.2ex] {11.72};
    \draw[|-|] (T33.north) -- (T43) node [rectangle,midway,fill=white,inner sep=1.2ex] {28.02};
    \draw[|-|] (T34.north) -- (T44) node [rectangle,midway,fill=white,inner sep=1.2ex] {1.62};
    \draw[|-|] (T35.north) -- (T45) node [rectangle,midway,fill=white,inner sep=1.2ex] {1.67};
    \draw[|-|] (T36.north) -- (T46) node [rectangle,midway,fill=white,inner sep=1.2ex] {185.21};
    \draw[|-|] (Txfer11.north) -- (Txfer21.south) node [rectangle,midway,fill=white,inner sep=1.2ex] {15.70};
    \draw[|-|] (Txfer12.north) -- (Txfer22.south) node [rectangle,midway,fill=white,inner sep=1.2ex] {1.66};
    
    \draw[|-|] (S11.north) -- (S21.south) node [rectangle,midway,fill=white,inner sep=1.2ex] {float};
    \draw[|-|] (S12.north) -- (S22.south) node [rectangle,midway,fill=white,inner sep=1.2ex] {17.11};
    \draw[|-|] (S13.north) -- (S23.south) node [rectangle,midway,fill=white,inner sep=1.2ex] {2.27};
    \draw[|-|] (S14.north) -- (S24.south) node [rectangle,midway,fill=white,inner sep=1.2ex] {197.11}; 
    \draw[|-|] (S31.north) -- (S41) node [rectangle,midway,fill=white,inner sep=1.2ex] {double}; 
    \draw[|-|] (S32.north) -- (S42) node [rectangle,midway,fill=white,inner sep=1.2ex] {20.61};
    \draw[|-|] (S33.north) -- (S43) node [rectangle,midway,fill=white,inner sep=1.2ex] {54.58};
    \draw[|-|] (S34.north) -- (S44) node [rectangle,midway,fill=white,inner sep=1.2ex] {2.10};
    \draw[|-|] (S35.north) -- (S45) node [rectangle,midway,fill=white,inner sep=1.2ex] {2.30};
    \draw[|-|] (S36.north) -- (S46) node [rectangle,midway,fill=white,inner sep=1.2ex] {198.17};
    \draw[|-|] (Sxfer11.north) -- (Sxfer21.south) node [rectangle,midway,fill=white,inner sep=1.2ex] {29.45};
    \draw[|-|] (Sxfer12.north) -- (Sxfer22.south) node [rectangle,midway,fill=white,inner sep=1.2ex] {2.07};
    
    \draw[|-|] (M11.north) -- (M21.south) node [rectangle,midway,fill=white,inner sep=1.2ex] {float};
    \draw[|-|] (M12.north) -- (M22.south) node [rectangle,midway,fill=white,inner sep=1.2ex] {81.57};
    \draw[|-|] (M13.north) -- (M23.south) node [rectangle,midway,fill=white,inner sep=1.2ex] {1.99};
    \draw[|-|] (M14.north) -- (M24.south) node [rectangle,midway,fill=white,inner sep=1.2ex] {203.93}; 
    \draw[|-|] (M31.north) -- (M41) node [rectangle,midway,fill=white,inner sep=1.2ex] {double}; 
    \draw[|-|] (M32.north) -- (M42) node [rectangle,midway,fill=white,inner sep=1.2ex] {117.80};
    \draw[|-|] (M33.north) -- (M43) node [rectangle,midway,fill=white,inner sep=1.2ex] {211.83};
    \draw[|-|] (M34.north) -- (M44) node [rectangle,midway,fill=white,inner sep=1.2ex] {3.80};
    \draw[|-|] (M35.north) -- (M45) node [rectangle,midway,fill=white,inner sep=1.2ex] {2.85};
    \draw[|-|] (M36.north) -- (M46) node [rectangle,midway,fill=white,inner sep=1.2ex] {204.81};
    \draw[|-|] (Mxfer11.north) -- (Mxfer21.south) node [rectangle,midway,fill=white,inner sep=1.2ex] {116.39};
    \draw[|-|] (Mxfer12.north) -- (Mxfer22.south) node [rectangle,midway,fill=white,inner sep=1.2ex] {3.80};
\end{tikzpicture}
\end{table*}

Table \ref{tab:cpu_baseline} reports the performance and energy usage for a single FPGA kernel with different numerical representations of each benchmark problem size. It should be highlighted that as we are focused on single kernel at this stage we only include the tiny, small, and medium problem sizes due to the fact that large and huge must run multi-kernel. This is because the maximum size of an individual buffer (per kernel) that can be transferred via XRT onto the Alveo is 4GB \cite{app_acceleration}. It can be seen that moving to fixed-point arithmetic or reduced precision is not a silver bullet, and in-fact the Xilinx HLS tooling is able to efficiently exploit the Alveo U280 for floating-point computation. In general, optimal performance is obtained when using floating-point representation and it is interesting that at the kernel execution level double precision tends to slightly outperform single and half precision. Furthermore it is difficult to predict the performance of the fixed-point configurations, for instance with 32-bit fixed-point there is a performance difference if one chooses 12 or 16 bits to represent the numbers above the decimal point. 

The average power draw in Watts is also interesting, where it can be seen that single precision floating-point arithmetic tends to draw more power than half precision for the small and medium benchmarks most likely due to the more complex core. Moreover, in comparison to fixed-point arithmetic, floating-point is competitive in terms of power draw, with the power draw difficult to predict for fixed-point arithmetic, with no real clear pattern between configurations. It can be seen that the main factor impacting performance and energy efficiency is whether the host is working with double or single precision numbers, as this significantly impacts data reordering and transfer times.

\section{Performance and energy profile}
\label{sec:performance}

Building on the work reported in Section \ref{sec:optimisation}, we replicated the number of kernels on the FPGA such that a subset of batches of paths is processed by each kernel concurrently. The insights gained by experimentation with numerical representation in Section \ref{sec:numeric} mean that we focused on double, single, and half precision floating-point arithmetic for our final overall experiments in this section.  

\begin{figure}[htb]
  \centering
  \includegraphics[scale=0.39]{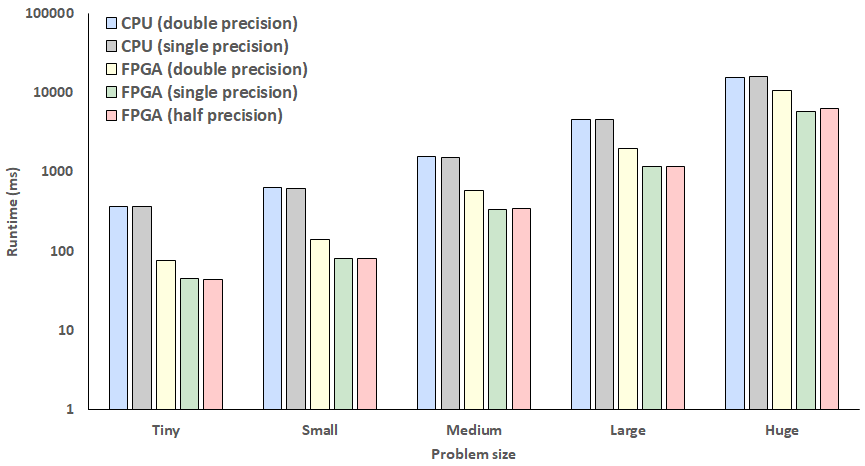}
  \caption{Runtime performance for FPGA kernels and CPU}
  \label{fig:total_performance}
\end{figure}

The resource utilisation reported in Table \ref{tab:resource_utilisation} illustrated that irrespective of the precision in use, the kernels require around 16\% of the FPGA's UltraRAM which limits the overall number of kernels to six. For single and half precision the UltraRAM utilisation constrains the number of kernels and therefore for problem sizes up to and including \emph{large} we build our kernels with 504 maximium timesteps only. This enables us to fit ten kernels onto the FPGA for such problem sizes, and this limit of ten is imposed by Vitis only making 32 ports available and each of our kernels requiring three ports. 1260 maximum timesteps is required for the huge problem size and consequently we are limited to six kernels for half and single precision. For double precision other resource constraints limit the number of kernels to six regardless of problem size. Figure \ref{fig:total_performance}, where the vertical axis is in log scale, reports the performance (in runtime) obtained by our FPGA kernel against the two 24-core Xeon Platinum CPUs for different problem sizes of the benchmark and floating-point precisions. It can be seen that irrespective of double or single, the CPU's performance is significantly worse than that obtained by the multiple FPGA kernels for all configurations, with single and half precision on the FPGA consistently fastest. There are two reasons why half and single are faster than double on the FPGA, firstly because of the difference in the number of kernels up to the \emph{large} problem size, and secondly because with double precision data transfer between the host and device is 64-bit, whereas with single and half it is 32-bit. Consequently there is twice the amount of data being reordered and transferred for double which results in additional overhead. For the huge problem size FPGA performance drops, not only because of the reduced number of kernels for single and half precision, but further more we must use the slower DDR-DRAM rather than HBM2 to fit the data.

\begin{figure}[htb]
  \centering
  \includegraphics[scale=0.39]{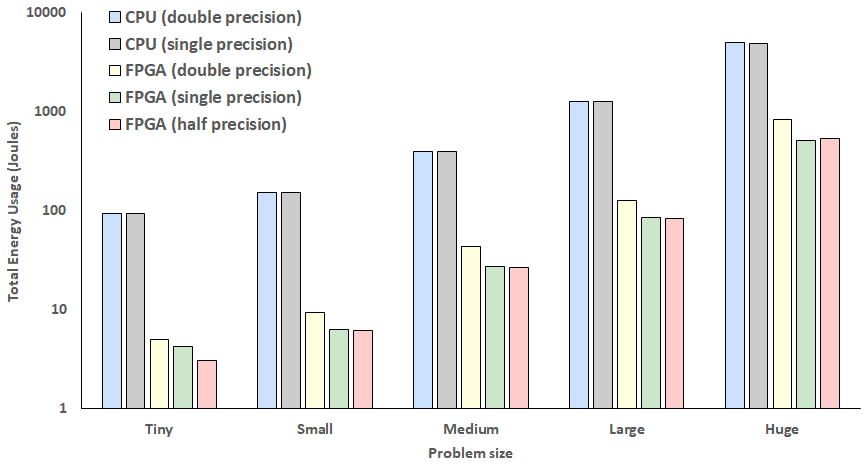}
  \caption{Total energy usage for FPGA kernels and CPU}
  \label{fig:total_energy}
\end{figure}

Figure \ref{fig:total_energy} reports the overall energy usage, in Joules, of our experiments, where the vertical axis is log scale. It can be seen that the two 24-core Xeon Platinum CPUs require the most energy which is from a combination of poorest performance and highest power draw. Single and half precision requires the least overall energy, and double precision on the FPGA is considerably more energy efficient than the CPU but worse than single and half precision. The reason for the increased energy requirement of double precision on the FPGA is a combination of the lower performance and increased power draw on both the FPGA and host data reordering. Energy for the FPGA increases significantly for the huge problem size because we are using DDR-DRAM in that configuration. 


\section{Conclusions and further work}
\label{sec:conclusions}
In this paper we have explored the role of FPGAs in delivering efficiency-driven computing for market risk analysis via the STAC-A2 Heston and Longstaff and Schwartz models on an Alveo U280 FPGA. Describing the algorithmic level dataflow optimisations that resulted in over 320 times increase in performance on the FPGA between the initial Von Neumann kernel and optimised dataflow algorithm, we then explored the role of different numerical representations and precision with the observation that floating-point arithmetic is highly competitive against fixed-point using the latest Xilinx Vitis toolchain and Alveo FPGA family for performance, power draw, energy efficiency, and resource utilisation. The major performance advantage at the single kernel level in moving to reduced precision was in reducing the overhead of data reordering on the host and data transfer via PCIe between the host and device. When moving to multiple FPGA kernels the smaller amount of resources required for single and half precision floating-point meant that more kernels could fit onto the chip, increasing concurrency.

Based upon insights gained from these explorations we then undertook detailed performance and energy comparisons for our benchmark kernel across different problem sizes, demonstrating that the FPGA is able to undertake the computation at significantly higher performance compared to the two 24-core Xeon Platinum Cascade Lake CPUs, and this combination of much higher performance and significantly reduced power draw has meant that the over all energy usage is very much less on the FPGA compared with the CPU, especially when the data fits into HBM2.

An important area of further work will be to explore streaming data to and from the FPGA rather than bulk copying all data to the device before execution begins. Currently data reordering and transfer accounts for up to a third of the runtime reported in Section \ref{sec:performance}, and a streaming approach would enable smaller chunks of data to be transferred before beginning kernel execution and to initiate transfers when a chunk has completed reordering on the host. However the Alveo U280 shell only supports AXI4 interface rather than AXIS which will complicate such an approach. Furthermore, we also plan to target the AI engines of Xilinx's next generation Versal architecture, where the chip contains up to 400 of these engines and each is a (single precision) floating-point or arbitrary precision fixed-point vectorised accelerator. Leveraging the AI engines could compliment the insights reported in Table \ref{tab:cpu_baseline}, for instance as 32-bit fixed-point delivers acceptable accuracy it would be interesting if the AI engines resulted in increased performance and reduced energy usage. 

We conclude that the result of this work not only demonstrates the clear benefit of leveraging an FPGA for efficiency-driven market risk analysis quantitative finance workloads, but furthermore the optimisation techniques described and lessons learnt from numerical experiments are applicable more widely to computational models on FPGAs.

\begin{acks}
The authors would like to thank STAC for access to the STAC-A2 benchmark and for their advice and assistance. We also acknowledge the ExCALIBUR H\&ES FPGA testbed and Xilinx XACC program for access to compute resource used in this work.
\end{acks}

\bibliographystyle{ACM-Reference-Format}
\bibliography{bib/bib-stac.bib}

\newpage
\clearpage

\appendix

\setcounter{table}{0}
\renewcommand{\thetable}{A\arabic{table}}

\noindent
\begin{minipage}{\textwidth}

\begin{center} 
  \captionof{table}{Overall multi-kernel performance and energy efficiency}
  \label{tab:perf_energy}
  \begin{tabular}{c|c|cc|cc|cc|cc|c}
    \toprule
    \makecell{\textbf{Problem}} 
    & \makecell{\textbf{dtype}} 
    & \multicolumn{2}{c|}{\textbf{Runtime (ms)}}
    & \multicolumn{2}{c|}{\textbf{Energy (J)}}
    & \multicolumn{2}{c|}{\textbf{Performance $\left(\text{s}^{-1}\right)$}}
    & \textbf{Eff\_B}
    & \textbf{Eff\_A}
    & \textbf{Eff\_A / Eff\_B}\\
    
    \makecell{\textbf{size}} 
    & 
    & CPU
    & FPGA
    & CPU
    & FPGA
    & CPU
    & FPGA
    & CPU
    & FPGA
    & Improvement\\
   \midrule
    Tiny     
             &  float      & 372.15 & 46.11 & 93.72 & 4.26               &2.687089 & 21.687270 & 0.02867145 & 5.09090835 & 177.56\\
             &  double     & 369.07 & 75.97 & 94.35 & 4.99               &2.709513 & 13.163091 & 0.02871768 & 2.63789393 & 91.86\\
    \midrule                                                                     
    Small    
             &  float      & 629.18 & 82.01 & 151.50 & 6.27              &1.589370 & 12.193635 & 0.01049089 & 1.94475836 & 185.38\\
             &  double     & 638.67 & 141.39 & 153.85 & 9.33             &1.565754 & 7.072636 & 0.01017715 & 0.75805316 & 74.49\\
    \midrule                                                                      
    Medium   
             &  float      & 1545.94 & 341.82 & 397.48 & 27.22           &0.646856 & 2.925516 & 0.00162739 & 0.10747672 & 66.04\\
             &  double     & 1551.90 & 592.79 & 393.34 & 43.89           &0.644371 & 1.686938 & 0.00163820 & 0.03843559 & 23.46\\
    \midrule                                                                     
    Large    
             &  float      & 4574.41 & 1178.14 & 1277.14 & 84.46         &0.218607 & 0.848796 & 0.00017117 & 0.01004968 & 58.71\\
             &  double     & 4558.11 & 1971.64 & 1266.59 & 125.41        &0.219389 & 0.507192 & 0.00017321 & 0.00404427 & 23.35\\
    \midrule                                                                      
    Huge     
             &  float      & 15825.42 & 5838.23 & 5011.16 & 510.49       &0.063189 & 0.171285 & 0.00001261 & 0.00033553 & 26.61\\
             &  double     & 15561.09 & 10644.86 & 4900.16 & 835.14      &0.064263 & 0.093942 & 0.00001311 & 0.00011249 & 8.58\\

 \bottomrule
\end{tabular}
\end{center}

\end{minipage}

\section{Performance Evaluation Document}

Table \ref{tab:perf_energy} provides data for questions 8 to 14. This table includes the overall evaluation of multi-kernel performance and energy efficiency for the baseline (CPU) and the accelerated (FPGA) system.


\subsection{Target system}


For the FPGA runs reported in this paper we use a Xilinx Alveo U280, running at the default clock frequency of 300MHz
. The FPGA card is hosted in a system with a 26-core Intel Xeon Platinum (Skylake) 8170 CPU and energy metrics gathered via XRT. All bitstreams are built using the Xilinx Vitis framework version 2021.2 which at the time of writing is the latest version. 

\subsection{Baseline system}


We compare the STAC-A2 benchmark specific hardware configured on FPGA against the optimised CPU version. All CPU runs are undertaken by threading using OpenMP across two 24-core Xeon Platinum (Cascade Lake) 8260M CPUs at 2.40GHz which are fitted into a single node of our test system and energy measured via RAPL. All CPU runs are executed across all 48 physical cores as this was found to be the optimal single-node CPU configuration.

\subsection{Target benchmark}


Major components of the STAC-A2 derivatives risk analysis benchmark which comprises market risk analysis workloads including monte carlo methods, the Heston stochastic volatility model, the Andersen Quadratic Exponential (QE) method and the Longstaff and Schwartz model for option pricing in quantitative finance.

\subsection{Performance metric}



Market risk analysis is a critical workload for trading floors and our chosen problem sizes reflect real-world workloads as defined in Table \ref{tab:prob_sizes_defined}. These problem sizes are used throughout this work in evaluating the CPU and FPGA benchmark implementations. While these workloads are executed multiple times on a daily basis, trading floors typically require results from computations within tight time frames. Therefore, for performance, we report the runtime (in ms) of these workloads for the chosen real-world problem sizes.

\newpage

\vspace*{6.05cm} 

\subsection{Energy measurement metric}


Energy consumption in Joules is measured for the whole FPGA card, and energy reported for the FPGA also includes energy required for data reordering on the host and transfer to/from the host. For the CPU baseline, we measure the CPU energy consumption in Joules for all 48 physical cores across two CPUs. 

\subsection{Experimental procedure to measure performance} 


All experiments undertaken report runtime from executing the Heston stochastic volatility model and path reduction in Longstaff and Schwartz. FPGA runtime combines execution, data transfer, and data reordering time. FPGA execution and data transfer time is gathered using OpenCL profiling information via \emph{getProfilingInfo} on the corresponding OpenCL events, which returns time at nanosecond resolution. Timing on the CPU is gathered using the \emph{clock\_gettime} call which returns time in nanosecond resolution. 


\subsection{Experimental procedure to measure power or energy consumption} 


We run the five classes of problem sizes. For each problem size on the baseline CPU, we measured the energy consumption across all 48 CPU cores via RAPL. For the FPGA, we measured the average power draw over the FPGA runtime via XRT. All reported results are averaged over five runs and total FPGA runtime and energy usage includes measurements of the kernel, data transfer and any required data reordering on the host.

\subsection{Other required metrics}
See Table \ref{tab:perf_energy} which reports the following metrics (questions 8 to 14) across our benchmark sizes and selected floating-point data types:

\begin{itemize}

\item \textbf{Performance result (target system)}
\item \textbf{Power or energy result (target system)}
\item \textbf{Efficiency of target system (Eff\_A)}
\item \textbf{Performance result (baseline)}
\item \textbf{Power or energy result (baseline)}
\item \textbf{Efficiency of baseline system (Eff\_B)}
\item \textbf{Efficiency improvement (Eff\_A / Eff\_B)} 

\end{itemize}


\end{document}